\documentclass[10pt]{article}
\usepackage{agop-arxiv}
\hypersetup{
  pdftitle={Explaining El Nino Forecasts with the Average Gradient Outer Product},
  pdfauthor={Yuan Hui, Dorian S. Abbot, and Robert J. Webber},
  pdfsubject={Explainable artificial intelligence for ENSO forecasting}
}
\begin{document}

\title{Explaining El Ni{\~n}o Forecasts with the Average Gradient Outer Product}

\authors{Yuan Hui\affil{1}, Dorian S. Abbot\affil{2}, and Robert J. Webber\affil{1}}

\affiliation{1}{Department of Mathematics, University of California San Diego}
\affiliation{2}{Department of the Geophysical Sciences, The University of Chicago}

\authoraddr{Robert J. Webber, Department of Mathematics, University of California San Diego, 9500 Gilman Drive, Mail Code 0112, La Jolla, CA 92093-0112, USA (rwebber@ucsd.edu)}

\authorrunninghead{Hui et al.}
\titlerunninghead{AGOP XAI for ENSO Forecasts}

\keypoints
{We introduce a new method based on the average gradient outer product (AGOP) to explain neural-network predictions in the physical sciences.}
{We define sensitivity, attribution, robustness, and coherence scores; AGOP XAI earns top scores in the latter three categories.}
{AGOP XAI reveals ocean precursors to extreme ENSO events and provides effective directions for steering the underlying physical Zebiak--Cane model.}

\maketitle

\begin{abstract}
An important and unresolved problem in the physical sciences is explaining the predictions made by neural networks.
Several explainable artificial intelligence (XAI) methods have been proposed to address this problem, including gradient XAI, Integrated Gradients, and GradientSHAP.
We evaluate the baseline XAI methods according to four scores:
\textbf{sensitivity} (XAI patterns strongly affect predictions),
\textbf{attribution} (XAI patterns reproduce the change in prediction relative to a baseline),
\textbf{robustness} (XAI patterns remain stable for nearby inputs), and
\textbf{coherence} (XAI patterns are spatially smooth).
We also introduce a new method---average gradient outer product (AGOP) XAI---that uses global gradient information to identify an important direction for a specific input.
We apply XAI to neural network predictions of the El Ni{\~n}o--Southern Oscillation (ENSO) based on data from the Zebiak--Cane model.

AGOP XAI achieves the highest attribution, robustness, and coherence scores in the architecture and lead-time comparisons reported here.
Its sensitivity is surpassed by gradient XAI, which is maximally sensitive by definition.
Beyond diagnosing neural-network behavior, AGOP XAI can generate candidate hypotheses about physical mechanisms.
The method highlights an equatorial thermocline-depth signal consistent with recharge oscillator physics, together with a southeastern-Pacific lobe that may be specific to the Zebiak--Cane model.
Finally, we test the physical relevance of AGOP using optimized perturbations that move the Zebiak--Cane model along AGOP explanation coordinates.
Such perturbations can suppress the selected extreme events or, from a near-neutral ensemble, generate strong El Ni{\~n}o or La Ni{\~n}a events 10 months later.

\begin{plainlanguagesummary}
Neural networks can make accurate predictions, but they do not identify which specific input patterns are driving their predictions.
Explainable artificial intelligence (XAI) is intended to provide the missing explanations.
Here, we evaluate established XAI methods and introduce AGOP XAI, which combines gradient information from many training samples.
We apply these methods to forecasts of extreme El Ni{\~n}o and La Ni{\~n}a events in a tropical Pacific climate model.
AGOP XAI earns the highest scores in three of our four tests, identifies ocean patterns consistent with a well-established theory of El Ni{\~n}o, and provides effective directions for steering the climate model itself.
\end{plainlanguagesummary}
\end{abstract}

\section{Introduction}
\label{sec:introduction}

Long-range prediction of the El Ni{\~n}o--Southern Oscillation (ENSO) provides a demanding test case for explaining neural networks in the physical sciences.
ENSO is a major source of year-to-year climate variability worldwide.
Current AI models can predict El Ni{\~n}o events at lead times of up to 1.5 years \cite{ham2019deep,liu2023explainable}, but it remains unclear why such long-range predictive accuracy is possible and whether AI models can improve our understanding of extreme El Ni{\~n}o dynamics.
We therefore seek an explainable artificial intelligence (XAI) method that identifies the specific input patterns responsible for individual forecasts.

To formulate this objective, we introduce a state vector $\bm{x} = (x_1, \ldots, x_d)$ that combines relevant geophysical fields such as sea surface temperature, zonal ocean current, and meridional ocean current at grid points across the spatial domain.
An AI model $f:\mathbb{R}^d\rightarrow\mathbb{R}$ uses this state vector to predict a quantitative outcome, such as the Ni{\~n}o-3 index, at some lead time in the future.
Given an input state $\bm{x}$, the goal of XAI is to justify each prediction $f(\bm{x})$ by producing a unit-norm \emph{explanation vector}
\begin{equation*}
    \bm{e}(\bm{x})
=
\bigl(e_1(\bm{x}),\ldots,e_d(\bm{x})\bigr),
\end{equation*}
which identifies an input pattern that drives the prediction.

Existing XAI methods may produce explanation vectors that lack robustness or physical meaning.
We propose that an effective explanation for a physical-science application should satisfy four criteria.
\begin{enumerate}
\item \textbf{Sensitivity.}
A small change in the explanation direction,
$\bm{x}\mapsto\bm{x}+\delta\bm{e}(\bm{x})$, produces a large first-order change in the prediction.

\item \textbf{Attribution.}
Relative to an average state $\bm{\mu}$, the explanation identifies the
part of the input anomaly $\bm{x}-\bm{\mu}$ responsible for the prediction
difference $f(\bm{x})-f(\bm{\mu})$.

\item \textbf{Robustness.}
The explanation remains stable for nearby input points.

\item \textbf{Coherence.}
The explanation varies smoothly across neighboring spatial grid points.
\end{enumerate}

Our first contribution is to convert these four criteria into quantitative
scores, each with an optimal value of $+1$.
The sensitivity score lies in $[0,1]$, while the attribution, robustness,
and coherence scores lie in $[-1,1]$.
Together, these scores provide a framework for systematically
comparing XAI methods.

Our second contribution is AGOP XAI, an explanation method based on the average gradient outer product.
AGOP XAI combines gradient information across the training data rather than relying on a single local gradient.
It uses this combined gradient information to provide an explanation specific to the input.

In experiments, we train neural networks to forecast ENSO from the output of the Zebiak--Cane model \cite{zebiak1987model}.
Then we apply AGOP XAI, gradients, Integrated Gradients, and
GradientSHAP to explain the neural network forecasts.
Across the architecture and lead-time comparisons reported here, AGOP XAI achieves the highest attribution, robustness, and coherence scores.
The gradient achieves the highest sensitivity score, as expected from its construction.

The AGOP explanations also reveal substantial differences among the neural-network architectures.
For the extreme El Ni{\~n}o prediction studied in detail, the CNN produces the most coherent AGOP explanation, while the MLP explanation contains small-scale irregularities and the ViT explanation contains conspicuous patchwise artifacts.

AGOP XAI can help identify candidate physical mechanisms for individual extreme events.
The CNN explanation for one extreme El Ni{\~n}o prediction at a 10-month lead contains an equatorial thermocline signal consistent with recharge oscillator theory \cite{jin1997equatorial,meinen2000warmwater}, together with a South-Pacific lobe not predicted by the simplest symmetric recharge picture.
The explanation for a strong La Ni{\~n}a prediction contains approximately sign-reversed thermocline patterns, consistent with the opposite phase of ENSO.
These results illustrate how XAI can generate physically interpretable hypotheses about extreme-event predictions.

Finally, we can use the CNN explanations to steer the Zebiak--Cane model.
Nudging the model along an extreme ten-month-ahead El Ni{\~n}o or La Ni{\~n}a explanation generally produces a strong event ten months later.
Conversely, nudging the model in the direction opposite the extreme El Ni{\~n}o or La Ni{\~n}a explanation can prevent a strong event from occurring.

The rest of the paper is organized as follows.
Section~\ref{sec:xai_methods} introduces the XAI methods and quantitative
evaluation scores.
Section~\ref{sec:methods} describes the data and neural networks used for ENSO
prediction.
Section~\ref{sec:results} presents the results and discussion, and
Section~\ref{sec:conclusions} concludes.

\section{XAI methods and quantitative scores} \label{sec:xai_methods}

In this section, we introduce three baseline XAI methods (Section~\ref{sec:context}) and illustrate their behavior through a simple example (Section~\ref{sec:example}).
We then define four quantitative scores for XAI evaluations (Section~\ref{sec:quantification}) and introduce AGOP XAI, which is designed to achieve strong performance across all four scores (Section~\ref{sec:agop}).

\subsection{Baseline XAI methods} \label{sec:context}

In the XAI literature, a method that generates an explanation vector for a single prediction is called a post-hoc local XAI method. It is post hoc because it is applied to an already trained model. It is local because it explains the prediction $f(\bm{x})$ at one particular input $\bm{x}=(x_1,\ldots,x_d)$. By contrast, a global XAI method identifies patterns in the model's behavior across many input points.
For each local method below, we normalize its nonzero output to obtain a unit-norm explanation vector $\bm{e}(\bm{x})$.

The simplest post-hoc local XAI technique \cite{baehrens2010how} returns the gradient of the prediction function with respect to the input data, $\nabla f(\bm{x}) = (\partial_1 f(\bm{x}), \ldots, \partial_d f(\bm{x}))$, where $\partial_i$ is the partial derivative with respect to the $i$th coordinate.
For a small input perturbation $\bm{\varepsilon}=(\varepsilon_1,\ldots,\varepsilon_d)$, the gradient gives the first-order approximation
\begin{equation*}
	f(\bm{x} + \bm{\varepsilon}) - f(\bm{x}) \approx \sum_{i=1}^d \partial_i f(\bm{x}) \varepsilon_i = \nabla f(\bm{x})^\top \bm{\varepsilon}.
\end{equation*}
Thus, the gradient explanation points in the input direction that produces the largest first-order change in the prediction.

The Integrated Gradients method \cite{sundararajan2017axiomatic} compares the prediction $f(\bm{x})$ with the prediction $f(\bm{\mu})$ at a reference state $\bm{\mu}=(\mu_1,\ldots,\mu_d)$, which we choose to be the average state.
By the fundamental theorem of calculus, the prediction difference can be expressed as
\begin{equation*}
    f(\bm{x}) - f(\bm{\mu}) = \sum_{i=1}^d \psi_i(\bm{x}).
\end{equation*}
Here, $\bm{\psi}(\bm{x})=(\psi_1(\bm{x}), \ldots, \psi_d(\bm{x}))$ is the Integrated Gradients attribution vector, with components
\begin{equation*}
     \psi_i(\bm{x}) = (x_i - \mu_i) \int_0^1 \partial_i f\bigl(r\bm{x} + (1-r) \bm{\mu}\bigr) \,\mathrm{d}r.
\end{equation*}
Using the element-wise product $\odot$, we can write this vector more compactly as
\begin{equation*}
     \bm{\psi}(\bm{x}) = (\bm{x} - \bm{\mu}) \odot \int_0^1 \nabla f\bigl(r\bm{x} + (1-r) \bm{\mu}\bigr) \,\mathrm{d}r.
\end{equation*}
Thus, the Integrated Gradients method multiplies the deviation from the reference state, $\bm{x} - \bm{\mu}$, by the average model gradient along a path from $\bm{\mu}$ to $\bm{x}$.

Several other post-hoc local XAI methods are closely related to input
gradients and Integrated Gradients.
For example, SmoothGrad averages gradients evaluated at randomly
perturbed versions of the input
\cite{smilkov2017smoothgradremovingnoiseadding}.
Gradient $\times$ Input takes the element-wise product of the
gradient and the input vector \cite{shrikumar2017learning}.
Finally, layer-wise relevance propagation
\cite{bach2015pixel,montavon2019layerwise,shrikumar2017learning}
propagates relevance backward through the network using layer-specific
redistribution rules, much like the chain rule used in differentiation
\cite{ancona2018towards}.
Here we focus on the simple gradient and Integrated Gradients
methods, which perform similarly to more complicated gradient-based methods in
geoscience applications \cite{mamalakis2022investigating,bommer2024finding}.

Our final post-hoc local XAI baseline is called Shapley additive explanations or SHAP \cite{lundberg2017unified}.
SHAP compares the prediction $f(\bm{x})$ with the average prediction over
a user-provided reference data set,
$\bm{z}_1, \ldots, \bm{z}_n$.
The prediction difference can be decomposed as
\begin{equation*}
    f(\bm{x})-\frac{1}{n}\sum_{i=1}^n f(\bm{z}_i)
    = \sum_{j=1}^d \phi_j(\bm{x}),
\end{equation*}
where each $\phi_j(\bm{x})$ is a Shapley value representing the
marginal contribution of variable $j$ to the prediction.
Several mathematical formulations of SHAP have been proposed
\cite{lundberg2017unified,bordt2023from}, together with practical
approximations designed to control the computational cost
\cite{erion2020learning,molnar2025interpretable}.
In this paper, we use the GradientSHAP approximation
\cite{erion2020learning}.
The attribution components are
\begin{equation*}
    \phi_j(\bm{x})
    =
    \frac{1}{n}\sum_{i=1}^n
    (x_j-z_{ij})
    \int_0^1
    \partial_j f\bigl(r\bm{x}+(1-r)\bm{z}_i\bigr)
    \,\mathrm{d}r.
\end{equation*}
Equivalently, using the element-wise product $\odot$, the GradientSHAP
attribution vector is
\begin{equation*}
    \bm{\phi}(\bm{x})
    =
    \frac{1}{n}\sum_{i=1}^n
    (\bm{x}-\bm{z}_i)\odot
    \int_0^1
    \nabla f\bigl(r\bm{x}+(1-r)\bm{z}_i\bigr)
    \,\mathrm{d}r.
\end{equation*}
Thus, GradientSHAP averages Integrated Gradients attributions computed
from multiple reference states.

Finally, several researchers \cite{mamalakis2022investigating,mamalakis2023carefully,bommer2024finding} distinguish between sensitivity and attribution methods.
Sensitivity methods seek input directions that produce a large first-order change in the prediction:
\begin{equation}
\label{eq:sensitivity}
    |f(\bm{x} + \delta \bm{e}(\bm{x})) - f(\bm{x})| \text{ is large for small } |\delta|.
\end{equation}
The gradient is the canonical sensitivity method.
In contrast, attribution methods seek to ``highlight those features in the input for which the deviation from the average state is important for the prediction'' \cite{mamalakis2023carefully}. Standard examples include Integrated Gradients and SHAP.

Traditional attribution methods introduce a reference state $\bm{\mu}$ and decompose the prediction difference $f(\bm{x}) - f(\bm{\mu})$ into separate contributions from different input coordinates.
We propose an alternative definition of an attribution as a component of the input anomaly, $\bm{x}-\bm{\mu}$, that approximately reproduces the prediction anomaly.
For any explanation direction $\bm{e}(\bm{x})$, the projection of the input anomaly onto that direction is
\begin{equation}
\label{eq:projection}
    \alpha\bm{e}(\bm{x}),
    \qquad \text{where } \alpha
    =
    \bm{e}(\bm{x})^\top(\bm{x}-\bm{\mu}).
\end{equation}
A successful attribution reproduces the prediction difference using only
this projected component and therefore satisfies
\begin{equation}
\label{eq:change}
    f\bigl(\bm{\mu}+\alpha\bm{e}(\bm{x})\bigr)-f(\bm{\mu})
    \approx
    f(\bm{x})-f(\bm{\mu}).
\end{equation}
The next subsection illustrates the distinction between standard
coordinate-wise attribution and our global, directional formulation.

\subsection{A simple example motivating a stronger notion of attribution} \label{sec:example}

\begin{figure}[t]
\centering
    \begin{tikzpicture}[
        scale=1.5,
        >=Latex,
        axis/.style={->, gray!70, thin},
        level/.style={gray!45, thin, dashed},
        grad/.style={->, very thick, blue!70!black},
        attr/.style={->, very thick, orange!85!black},
        point/.style={circle, fill=blue!70!black, inner sep=2.2pt},
        labelbox/.style={
            fill=white,
            rounded corners=2pt,
            inner sep=3pt,
            font=\small
        }
    ]

    \draw[axis] (-2.25,0) -- (2.85,0)
        node[right] {$x_1$};
    \draw[axis] (0,-1.55) -- (0,1.75)
        node[above] {$x_2$};

    \begin{scope}
        \clip (-2.25,-1.5) rectangle (2.85,1.65);

        \foreach \c in {-2,0,2} {
            \draw[level]
                (-2.1,{-2.1-\c}) -- (2.4,{2.4-\c});
        }
    \end{scope}

    \node[gray!65, font=\scriptsize, rotate=45]
        at (-1.28,0.52) {$f(\bm{x})=-2$};
    \node[gray!65, font=\scriptsize, rotate=45]
        at (-0.82,-1.02) {$f(\bm{x})=0$};
    \node[gray!65, font=\scriptsize, rotate=45]
        at (1.62,-0.58) {$f(\bm{x})=2$};

    \draw[grad] (0,0) -- (1,-1);
    \node[labelbox, anchor=south west, align=left]
        at (0.72,-1.77)
        {Input and gradient\\point here.};

    \draw[attr] (0,0) -- (1,1);
    \node[labelbox, anchor=south west, align=left]
        at (1.08,0.78)
        {Attribution methods\\point here.};

    \draw[gray!70, thin]
        (0.18,-0.18) -- (0.36,0) -- (0.18,0.18);

    \end{tikzpicture}
    \caption{For the linear model $f(\bm{x})=x_1-x_2$ at $\bm{x}=(1,-1)$, the input and gradient point along $(1,-1)$, whereas Integrated Gradients and GradientSHAP return the orthogonal attribution vector $(1,1)$.}
    \label{fig:linear-xai-limitation}
\end{figure}
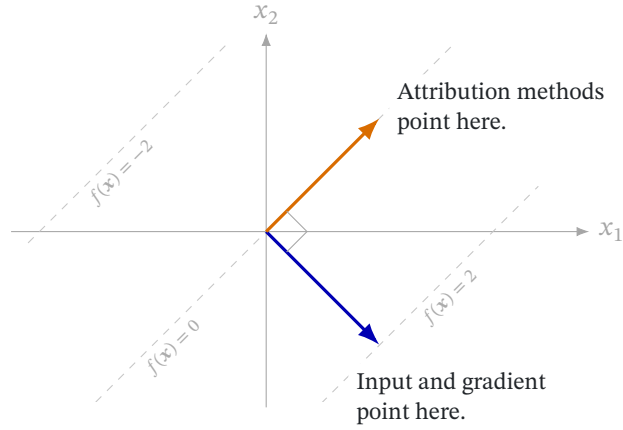

We illustrate the baseline XAI methods using the linear model
\begin{equation*}
    f(\bm{x}) = x_1-x_2,
\end{equation*}
at the input $\bm{x}=(1,-1)$, which leads to a prediction $f(\bm{x})=2$.
The input vector and level sets of $f$ are shown in
Figure~\ref{fig:linear-xai-limitation}.

In this setting, the gradient XAI method produces a vector
\begin{equation*}
    \nabla f(\bm{x}) = (1, -1).
\end{equation*}
After normalization,
this gradient explanation achieves perfect sensitivity and attribution for any baseline, as defined by equations~\eqref{eq:sensitivity}--\eqref{eq:change}.
The explanation satisfies an even stronger property: the model output can be recovered by projecting every input point onto the explanation direction, while the orthogonal component has no effect on the prediction.

For comparison, we compute Integrated Gradients using the baseline $\bm{\mu} = (0, 0)$.
Since the gradient is constant,
\begin{align*}
    \bm{\psi}(\bm{x}) 
    &= (\bm{x}-\bm{\mu}) \odot \int_0^1 \nabla f\bigl(r \bm{x} + (1-r)  \bm{\mu} \bigr) \,\mathrm{d}r \\
    &= \begin{bmatrix} 1 \\ -1 \end{bmatrix} \odot \begin{bmatrix} 1 \\ -1 \end{bmatrix} = \begin{bmatrix} 1 \\ 1 \end{bmatrix}.
\end{align*}
Similarly, we compute the GradientSHAP attribution vector given a mean-zero baseline data set $\bm{z}_1, \ldots, \bm{z}_n$.
Since the gradient is constant and the reference data set has mean zero,
\begin{align*}
    \bm{\phi}(\bm{x})
    &= \frac{1}{n} \sum_{i=1}^n (\bm{x} - \bm{z}_i) \odot \int_0^1 \nabla f\bigl( r \bm{x} + (1-r) \bm{z}_i \bigr)\,\mathrm{d}r \\
    &= \begin{bmatrix} 1 \\ -1 \end{bmatrix} \odot \begin{bmatrix} 1 \\ -1 \end{bmatrix} = \begin{bmatrix} 1 \\ 1 \end{bmatrix}.
\end{align*}
Thus, Integrated Gradients and GradientSHAP identify the same direction, $(1, 1)$.
This direction has zero sensitivity because
\begin{equation*}
    f(\bm{x}+\delta \bm{\psi}(\bm{x})) = f(\bm{x}),
    \qquad \text{for every } \delta.
\end{equation*}
This direction also fails to reproduce the prediction relative to the baseline.

In this example, Integrated Gradients and GradientSHAP succeed under the standard coordinate-wise interpretation of attribution: each coordinate contributes $+1$ to the prediction difference, the first through $1\times 1$ and the second through $(-1)\times(-1)$. However, the methods fail our stronger requirement that the explanation identifies a coordinated input pattern that reproduces the prediction.
The resulting direction $(1,1)$ has no effect on the prediction and is therefore not useful for scientists seeking to influence the predicted outcome.
This limitation echoes a recent analysis \cite{bilodeau2024impossibility} showing that attribution methods may identify relevant variables without specifying how changes in those variables affect model behavior.

\subsection{How we will quantify XAI success} \label{sec:quantification}

\citeA{bommer2024finding} recently provided a systematic assessment of XAI in geoscience.
The authors considered five desired properties: robustness, faithfulness, complexity, localization, and randomization.
Robustness means that small perturbations to the input produce only small changes in the explanation.
Faithfulness means that the prediction changes after perturbing or masking features identified as important.
Complexity favors concise explanations containing a small number of strong features.
Localization measures the agreement between the explanation and a predefined region of interest.
Finally, randomization means that the explanation changes when the model parameters are randomly perturbed or a different output class is explained.

Here, we develop an alternative XAI evaluation framework based on \textbf{sensitivity}, \textbf{attribution},
\textbf{robustness}, and \textbf{coherence}.
Each score has an optimal value of $+1$.
Attribution, robustness, and coherence lie in $[-1,1]$, while sensitivity lies in $[0,1]$.
We also identify ways in which our framework differs from that of \citeA{bommer2024finding}.

\textit{The model prediction should be sensitive to the patterns identified by XAI.} Many researchers argue
that XAI should identify patterns that strongly influence the model prediction
\cite{mamalakis2022investigating,bommer2024finding}.
We therefore define the sensitivity score by the
limiting ratio
\begin{equation*} 
    S_s(\bm{x}) = \lim_{\delta\rightarrow 0} \frac{\left|f\bigl(\bm{x}+\delta\bm{e}
    (\bm{x})\bigr)-f(\bm{x})\right|} {\lVert\delta\nabla f(\bm{x})\rVert},
\end{equation*}
where we assume that $f$ has a nonzero gradient at $\bm{x}$.
Since $\bm{e}(\bm{x})$ has unit norm, this is equivalent to
\begin{equation*} 
    S_s(\bm{x}) =
    \frac{\left|\nabla f(\bm{x})^\top\bm{e}(\bm{x})\right|} {\lVert\nabla f(\bm{x})\rVert}.
\end{equation*}
The score
lies in $[0,1]$. A value of $1$ means that the explanation is a direction of maximal first-order change, while a value of $0$ means that perturbing the input along the explanation has no first-order effect.

Our criterion differs from the faithfulness property of \citeA{bommer2024finding}. Rather than masking a few coordinates with the largest attribution magnitudes, we perturb the input according to coordinated large-scale spatial patterns.

\textit{The result of interest can be reproduced by the patterns XAI identifies.}
Many researchers agree that attribution is important \cite{mamalakis2022investigating,bommer2024finding}, and we use the ideal-attribution criterion discussed in Section~\ref{sec:context}.
We fix an average state $\bm{\mu}$ and determine how much the explanation overlaps with $\bm{x}-\bm{\mu}$:
\begin{equation*}
\alpha
=
\bm{e}(\bm{x})^\top
\bigl(\bm{x}-\bm{\mu}\bigr).
\end{equation*}
For inputs satisfying $f(\bm{x})\neq f(\bm{\mu})$, we compute the ratio
\begin{equation*}
R
=
\frac{
f\bigl(\bm{\mu}+\alpha\bm{e}(\bm{x})\bigr)-f(\bm{\mu})
}{
f(\bm{x})-f(\bm{\mu})
}.
\end{equation*}
We define the attribution score as
\begin{equation*}
S_a(\bm{x})
=
\begin{cases}
R, & R\in[-1,1], \\
1/R, & R>1 \text{ or } R<-1.
\end{cases}
\end{equation*}
A positive attribution score means that the explanation produces a change in the correct direction, and a higher magnitude means that the explanation more accurately reproduces the magnitude of the change in prediction.
The value $S_a(\bm{x})=1$ denotes exact reproduction.
The score treats equal-factor overestimates and underestimates symmetrically; for example, $R=2$ and $R=1/2$ both give $S_a(\bm{x})=1/2$.
The attribution score has no direct analogue in the framework of \citeA{bommer2024finding}.

\textit{XAI results should be robust to small perturbations of the input.} An explanation should reflect stable features of the model.
Consequently, sufficiently small input perturbations should produce only small changes in the explanation.
We define the robustness score as the average signed overlap between the original explanation vector and the explanation vectors obtained from perturbed inputs,
\begin{equation*}
    S_r(\bm{x}) = \frac{1}{k} \sum_{i=1}^k \bm{e}(\bm{x})^\top \bm{e}(\bm{z}_i).
\end{equation*}
Here, $\bm{z}_1,\ldots,\bm{z}_k\in\mathbb{R}^d$ are the $1\%$ of training inputs closest to $\bm{x}$ in standardized Euclidean distance.
Because the explanation vectors have unit norm, the
robustness score $S_r(\bm{x})$ lies in the interval $[-1,1]$, with higher values indicating greater robustness.

\citeA{bommer2024finding} defined a robustness score based on independent mean-zero perturbations of $\bm{x}$.
This has the disadvantage of creating unphysical data points.

\begin{figure}[t]
\centering
\publishedfit{%
\begin{tikzpicture}[
    x=1cm,
    y=1cm,
    edge/.style={gray!58, thin},
    orangeedge/.style={orange!85!black, line width=1.15pt},
    bottomtext/.style={align=center, font=\small},
    dot/.style={
        circle,
        draw=gray!65,
        line width=0.35pt,
        minimum size=6.3mm,
        inner sep=0pt
    },
    plusdot/.style={
        dot,
        draw=blue!75!black,
        fill=blue!75!black,
        text=white,
        font=\scriptsize\bfseries
    },
    minusdot/.style={
        dot,
        draw=orange!90!black,
        fill=orange!90!black,
        text=white,
        font=\scriptsize\bfseries
    }
]



\begin{scope}[shift={(-5.6,0)}]

    \foreach \yy in {-0.6,0.6,1.8,3.0} {
        \draw[edge] (-1.8,\yy) -- (1.8,\yy);
    }
    \foreach \xx in {-1.8,-0.6,0.6,1.8} {
        \draw[edge] (\xx,-0.6) -- (\xx,3.0);
    }

    \draw[orangeedge] (-0.6,-0.6) -- (0.6,-0.6);

    \foreach \xx in {-1.8,-0.6,0.6,1.8} {
        \foreach \yy in {-0.6,0.6,1.8,3.0} {
            \node[dot, fill=gray!10] at (\xx,\yy) {};
        }
    }

    \node[dot, draw=orange!85!black, fill=orange!85!black] at (-0.6,-0.6) {};
    \node[dot, draw=orange!85!black, fill=orange!85!black] at ( 0.6,-0.6) {};

    \node[bottomtext] at (0,-1.75)
        {Adjacent same-field grid points\\guarantee $\bm{A}(i,j)=1$.};

\end{scope}


\begin{scope}[shift={(0,0)}]

    \foreach \yy in {-0.6,0.6,1.8,3.0} {
        \draw[edge] (-1.8,\yy) -- (1.8,\yy);
    }
    \foreach \xx in {-1.8,-0.6,0.6,1.8} {
        \draw[edge] (\xx,-0.6) -- (\xx,3.0);
    }

    \foreach \yy in {-0.6,0.6,1.8,3.0} {
        \node[dot, fill=blue!10!white] at (-1.8,\yy) {};
        \node[dot, fill=blue!28!white] at (-0.6,\yy) {};
        \node[dot, fill=blue!52!white] at ( 0.6,\yy) {};
        \node[dot, fill=blue!82!white] at ( 1.8,\yy) {};
    }

    \node[bottomtext] at (0,-1.75)
        {Smooth patterns yield\\high coherence: $S_c(\bm{x}) \approx +1$.};

\end{scope}


\begin{scope}[shift={(5.6,0)}]

    \foreach \yy in {-0.6,0.6,1.8,3.0} {
        \draw[edge] (-1.8,\yy) -- (1.8,\yy);
    }
    \foreach \xx in {-1.8,-0.6,0.6,1.8} {
        \draw[edge] (\xx,-0.6) -- (\xx,3.0);
    }

    \node[plusdot]  at (-1.8, 3.0) {$+$};
    \node[minusdot] at (-0.6, 3.0) {$-$};
    \node[plusdot]  at ( 0.6, 3.0) {$+$};
    \node[minusdot] at ( 1.8, 3.0) {$-$};

    \node[minusdot] at (-1.8, 1.8) {$-$};
    \node[plusdot]  at (-0.6, 1.8) {$+$};
    \node[minusdot] at ( 0.6, 1.8) {$-$};
    \node[plusdot]  at ( 1.8, 1.8) {$+$};

    \node[plusdot]  at (-1.8, 0.6) {$+$};
    \node[minusdot] at (-0.6, 0.6) {$-$};
    \node[plusdot]  at ( 0.6, 0.6) {$+$};
    \node[minusdot] at ( 1.8, 0.6) {$-$};

    \node[minusdot] at (-1.8,-0.6) {$-$};
    \node[plusdot]  at (-0.6,-0.6) {$+$};
    \node[minusdot] at ( 0.6,-0.6) {$-$};
    \node[plusdot]  at ( 1.8,-0.6) {$+$};

    \node[bottomtext] at (0,-1.75)
        {Alternating patterns yield\\lowest coherence: $S_c(\bm{x}) = -1$.};

\end{scope}

\end{tikzpicture}%
}
\caption{
Illustration of spatial coherence.
Left: the adjacency graph connects coordinates from the same field at neighboring grid points.
Center: a smooth pattern has high coherence.
Right: an alternating checkerboard attains the minimum score, $S_c(\bm{x})=-1$.
}
\label{fig:spatial-coherence}
\end{figure}
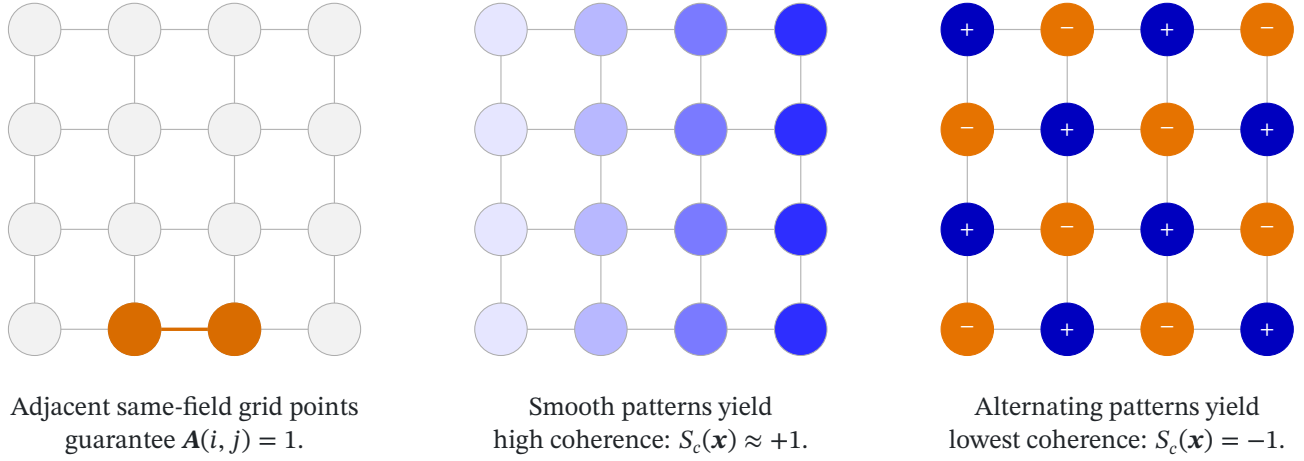

\textit{The patterns XAI identifies should be spatially smooth.}
Physical scientists often search for explanations linked to the partial differential equations describing physical processes.
Barring a few exceptions, such as shocks in a fluid, these explanations should involve smooth spatial patterns of fields and their gradients.
We quantify spatial smoothness using a standard graph-theoretic approach.
We define the graph adjacency matrix $\bm{A}\in \{0,1\}^{d\times d}$ by setting $\bm{A}(i,j)=1$ when variables $i$ and $j$ represent the same field at adjacent grid points and $\bm{A}(i,j)=0$ otherwise.
We define the degree matrix $\bm{D}\in\mathbb{R}^{d\times d}$ to have diagonal entries
\begin{equation*}
\bm{D}(i,i)=\sum_{j=1}^d \bm{A}(i,j),
\end{equation*}
and zero off-diagonal entries.
Assuming $\bm{e}(\bm{x})^\top \bm{D} \bm{e}(\bm{x}) \neq 0$, the coherence score is
\begin{equation*}
S_c(\bm{x})
=
\frac{
\bm{e}(\bm{x})^\top\bm{A}\bm{e}(\bm{x})
}{
\bm{e}(\bm{x})^\top\bm{D}\bm{e}(\bm{x})
}.
\end{equation*}
This score lies in $[-1,1]$.
The value $S_c(\bm{x})=+1$ corresponds to an explanation that is constant within each field, while
$S_c(\bm{x})=-1$ corresponds to an alternating pattern illustrated in Figure~\ref{fig:spatial-coherence}.

Our coherence score differs from the sparsity-based complexity score of \citeA{bommer2024finding}.
Sparsity ignores spatial relationships, whereas we regard spatial coherence as essential for physical-science explanations involving partial differential equations.

\subsection{Input-specific explanations from the average gradient outer product (AGOP)} \label{sec:agop}

This subsection introduces an XAI technique designed to balance the four criteria of \textbf{sensitivity}, \textbf{attribution}, \textbf{robustness}, and \textbf{coherence}.
A natural starting point is the local gradient, $\nabla f(\bm{x})$, which identifies the direction of maximal first-order sensitivity.
For a linear model, the normalized gradient also achieves a perfect attribution score.
However, for the nonlinear AI models commonly used in the physical sciences, the local gradient may fail to reproduce the prediction from a distant reference state.
Moreover, direct gradient explanations can be spatially noisy and can change sharply under small input perturbations \cite{mamalakis2022investigating}.

Rather than averaging the gradients themselves, which could cancel
gradients with opposite signs, we average their outer products.
Given a user-provided reference data set
$\bm{z}_1,\ldots,\bm{z}_n\in\mathbb{R}^d$, we define the average gradient outer product (AGOP),
\begin{equation*}
    \bm{M}
    =
    \frac{1}{n}
    \sum_{j=1}^n
    \nabla f(\bm{z}_j)
    \nabla f(\bm{z}_j)^\top.
\end{equation*}
The AGOP is positive semidefinite, and for every unit vector $\bm v$,
\begin{equation*}
    \bm v^\top\bm M\bm v
    =
    \frac{1}{n}
    \sum_{j=1}^n
    \left|
        \nabla f(\bm z_j)^\top\bm v
    \right|^2.
\end{equation*}
Consequently, its leading eigenvectors identify directions with large
average squared directional derivatives over the reference data.

The AGOP is a sample-based version of the gradient second-moment matrix, which has been used for dimension reduction, sensitivity analysis, and neural feature learning
\cite{xia2002adaptive,constantine2014active,zahm2022certified,
radhakrishnan2024mechanism}.
Traditional applications primarily extract its leading eigenvectors or leading eigenspace to identify sensitive input directions.
Recent work has begun to exploit more of the matrix.
In particular, recursive feature machines use the AGOP to rescale inputs and iteratively learn task-specific features
\cite{radhakrishnan2024mechanism}.
Concurrent work has also proposed local XAI methods that combine the AGOP with the gradient at the input \cite{katakam2026agop,katakam2026agopixg}.

Here, we use the full AGOP in a new way.
We apply $\bm{M}^{1/2}$ to the centered input to produce an explanation,
\begin{equation*}
    \bm{e}(\bm{x})
    =
    \frac{\bm{M}^{1/2}(\bm{x}-\bm{\mu})}{\lVert\bm{M}^{1/2}(\bm{x}-\bm{\mu}) \rVert},
    \qquad \text{where } \bm{\mu}=\bm{0} \text{ in our experiments}.
\end{equation*}
This explanation is well-defined when $\lVert\bm M^{1/2}(\bm x-\bm\mu) \rVert \neq 0$.
Here, $\bm{0}$ corresponds to the mean of the full 10,000-year training block in standardized coordinates.
To our knowledge, the square-root matrix--vector construction has not previously been used for XAI.

To distinguish AGOP XAI from a fixed global eigendirection,
write
\begin{equation*}
    \bm M
    =
    \sum_{k=1}^d
    \lambda_k\bm v_k\bm v_k^\top.
\end{equation*}
Then
\begin{equation*}
    \bm M^{1/2}(\bm x-\bm\mu)
    =
    \sum_{k=1}^d
    \sqrt{\lambda_k}\,
    \bm v_k
    \bm v_k^\top(\bm x-\bm\mu).
\end{equation*}
Thus, AGOP XAI combines the globally important directions according to
their presence in the input anomaly $\bm x-\bm\mu$.
Unlike the leading eigenvectors of $\bm M$, the resulting explanation
changes with the input and has a natural orientation determined by the input.

AGOP XAI has useful structural properties.
\begin{itemize}
\item
If $\bm M$ has rank one, then all gradients at the reference data points
lie in the span of a single vector $\bm v_1$.
In this case,
\begin{equation*}
    \bm e(\bm x)
    =
    \bm v_1
    \operatorname{sign}
    \bigl(\bm v_1^\top(\bm x-\bm\mu)\bigr).
\end{equation*}
Thus, AGOP XAI returns the unique important direction, with its orientation
determined by the input anomaly.
The method therefore exactly recovers the relevant direction for linear
and single-index models.
\item
The method behaves consistently under rotations and reflections of the
coordinate system.
If we rotate or reflect every input field, then the resulting explanation
is rotated or reflected in exactly the same way.
Thus, the explanation does not depend on an arbitrary choice of coordinate
orientation.

\item
If $\bm M$ is diagonal, the $i$th entry of
$\bm M^{1/2}(\bm x-\bm\mu)$ is
\begin{equation*}
    \underbrace{\vphantom{\left(\sum_j^n\right)}(x_i-\mu_i)}_{\text{input anomaly}}
    \underbrace{
    \left(
        \frac{1}{n}
        \sum_{j=1}^n
        \bigl[
            \partial_i f(\bm z_j)
        \bigr]^2
    \right)^{1/2}
    }_{\text{root-mean-squared derivative}}.
\end{equation*}
By comparison, the Integrated Gradients attribution uses
\begin{equation*}
    \underbrace{\vphantom{\int_0^1}(x_i-\mu_i)}_{\text{input anomaly}}
    \underbrace{\int_0^1
    \partial_i f
    \bigl(r\bm x+(1-r)\bm\mu\bigr)
    \,\mathrm{d}r}_{\text{signed average derivative}}.
\end{equation*}
Thus, AGOP XAI uses the root-mean-squared derivative over the reference data set in place of the signed average derivative
along a single path.
This example explains why the exponent $1/2$ is appropriate.
Taking the square root converts the mean square derivative $M_{ii}$ into the natural sensitivity scale.
\end{itemize}

In the application we introduce below, the input dimension is relatively small, $d=2{,}162$, so we form the AGOP directly and
compute its eigendecomposition.
In higher-dimensional applications, the full AGOP can be replaced by a low-rank approximation using either a deterministic Frequent Directions sketch of the gradient matrix in a one-pass setting \cite{ghashami2016frequent} or a randomized Nystr\"om block Krylov approximation of the positive-semidefinite AGOP operator when multiple passes are practical \cite{tropp2023randomizedalgorithmslowrankmatrix}.

\section{AI methods for ENSO prediction} \label{sec:methods}

\begin{figure}[t]
    \centering
    \includegraphics[width=\linewidth]{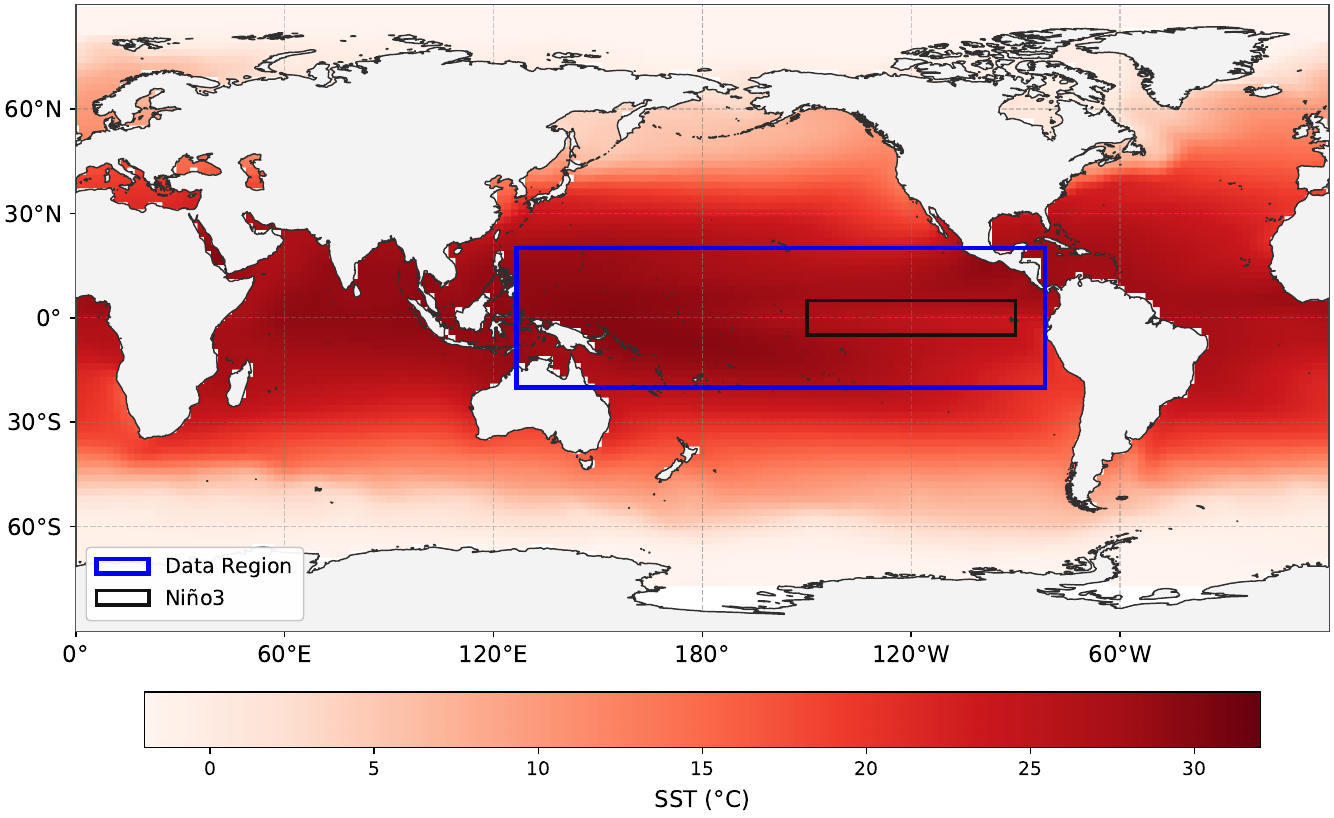}
    \caption{
    The black box marks the Ni{\~n}o-3 region, and the blue box marks the ZC input domain ($20^\circ$S--$20^\circ$N, $126.5625^\circ$E--$81.5625^\circ$W).
    We use SST anomaly, thermocline-depth anomaly, and zonal and meridional depth-averaged ocean currents to forecast the Ni{\~n}o-3 index.
    The background is the climatological mean SST from NOAA ERSSTv5 \protect\cite{NOAA_ERSSTv5}.
    }
    \label{fig:map}
\end{figure}

We study neural-network forecasts of the El Ni{\~n}o--Southern Oscillation (ENSO).
ENSO arises from interactions between the tropical Pacific Ocean and atmosphere and affects droughts and flooding around the world.
The recharge oscillator \cite{jin1997equatorial,meinen2000warmwater} provides a physical framework for ENSO predictability.
It describes how wind-stress anomalies recharge and discharge equatorial warm-water volume, altering the thermocline depth and influencing subsequent transitions between El Ni{\~n}o and La Ni{\~n}a.

Our specific task is to forecast the Ni{\~n}o-3 index using data generated from the Zebiak--Cane model \cite{zebiak1987model}.
The Ni{\~n}o-3 index is defined as the average sea-surface temperature (SST) anomaly over the eastern equatorial Pacific ($5^\circ$S--$5^\circ$N, $150^\circ$W--$90^\circ$W), as shown in Figure~\ref{fig:map}.
We compare three neural-network architectures: a multilayer perceptron (MLP), a convolutional neural network (CNN), and a vision transformer (ViT).

\subsection{Zebiak--Cane model}

The ZC model is an intermediate-complexity coupled ocean--atmosphere model of the tropical Pacific.
The atmosphere is a steady, single-layer model governed by the linear shallow-water equations.
The ocean consists of a shallow, warm, advecting upper layer separated from a motionless deep layer by an interface.
The interface between the upper and lower layers represents the thermocline, the region of strong vertical temperature gradients in the ocean.

The ZC model has been used extensively to study ENSO dynamics and
predictability \cite{zebiak1987model,jin1997equatorial}.
A central feature of the model is that it builds up equatorial upper-ocean heat content before warm events and discharges that heat during those events.
Indeed, the recharge oscillator can be derived as a reduction of Zebiak--Cane-type dynamics, with wind stress and zonal-mean equatorial thermocline depth governing the slow evolution of ENSO \cite{jin1997equatorial}.
The ZC model therefore lets us test whether XAI methods recover the large-scale thermocline structures associated with ENSO evolution.

We train, validate, and test the neural networks using 12,000 years of ZC output retained after a 100-year spinup.
We use the tropical Pacific domain whose cell edges span $20^\circ$S--$20^\circ$N and $126.5625^\circ$E--$81.5625^\circ$W, as shown in Figure~\ref{fig:map}.
Our neural-network input comprises four ZC model fields at each grid point: sea surface temperature anomaly, thermocline-depth anomaly, and zonal and meridional depth-averaged ocean currents.
We represent each field on a $20 \times 27$ latitude--longitude grid, giving 540 spatial grid points per field.
Then we append two scalar coordinates encoding the sine and cosine of annual phase, giving a total input dimension of $d=4\times20\times27+2=2{,}162$.

\subsection{Data processing}

We split the 12,000-year time series chronologically, using the first 10,000 years for training, the next 1,000 years for validation, and the remaining 1,000 years for testing.
We compute the mean and standard deviation of each of the 2,162 features over the training set, and we standardize the training, validation, and testing inputs using these training-set statistics.
We use the standardized inputs to train the models, compute the explanations, evaluate the XAI scores, and produce the visualizations.

The Ni{\~n}o-3 response variable is defined by spatially averaging the SST anomaly field; it is not standardized and remains in degrees Celsius.
The Ni{\~n}o-3 index is not constrained to have zero mean or median, because the underlying SST anomaly field for the ZC model is defined relative to a prescribed, annually repeating climatological background \cite{zebiak1987model}.
Across our 12,000-year simulation, the Ni{\~n}o-3 index is warm-skewed: its median is $+0.042^\circ\mathrm{C}$, whereas its mean is $+0.353^\circ\mathrm{C}$.

The simulation contains three time points per month.
For a lead time of $k$ months, we train the neural networks to predict the Ni{\~n}o-3 index $y_{t+3k}$ from $\bm{x}_t$.
We train a separate model at each lead time, and many results below use a 10-month lead.

\subsection{AI prediction models}

We train three neural networks on the ZC model output: a multilayer perceptron (MLP), a convolutional neural network (CNN), and a vision transformer (ViT). 
We choose compact configurations with comparable numbers of trainable parameters, ranging from 211,777 to 236,601.

MLPs provide a traditional fully connected baseline.
Our MLP consists of three fully connected hidden layers of width 100 with ReLU activations, followed by a linear scalar-output layer. All linear layers include bias terms, and the MLP has 236,601 trainable parameters.

CNNs have previously been used to predict El Ni{\~n}o events \cite{ham2019deep}.
Our CNN passes the four spatial input fields through two convolutional blocks.
Each block contains two $3 \times 3$ convolutional layers with same padding, 50 output channels, and ReLU activations, followed by average pooling. The first pooling operation uses nonoverlapping $2\times3$ windows, reducing the $20\times27$ grid to $10\times9$. The second uses nonoverlapping $2 \times 1$ windows, producing a $5 \times 9$ grid.
We flatten the resulting $50 \times 5 \times 9$ tensor, append the two annual-phase coordinates, and pass the resulting 2,252-dimensional vector through two fully connected hidden layers of width 64 with ReLU activations, followed by a linear scalar-output layer. All convolutional and fully connected layers include bias terms.
The CNN uses neither batch normalization nor dropout and has 217,917 trainable parameters.

Transformers have more recently been applied to weather forecasting \cite{bi2023accurate}.
Our ViT divides the spatial input into 90 nonoverlapping $2\times3$ patches.
Each patch contains four spatial fields and is therefore represented by $4\times2\times3=24$ values, which we map to a 64-dimensional embedding.
We add a distinct learnable 64-dimensional positional embedding to each patch.
The ViT contains four pre-normalized transformer encoder blocks with embedding dimension 64, four attention heads, and feed-forward width 256. Each block uses residual connections and a ReLU activation in its feed-forward sublayer.
After a final layer normalization, we average the 90 patch embeddings, append the two annual-phase coordinates, and pass the result through a fully connected hidden layer of width 64 with a ReLU activation and a linear output layer. The ViT uses no dropout and has 211,777 trainable parameters.

\begin{figure*}[t]
   \centering
    \includegraphics[width=\textwidth]{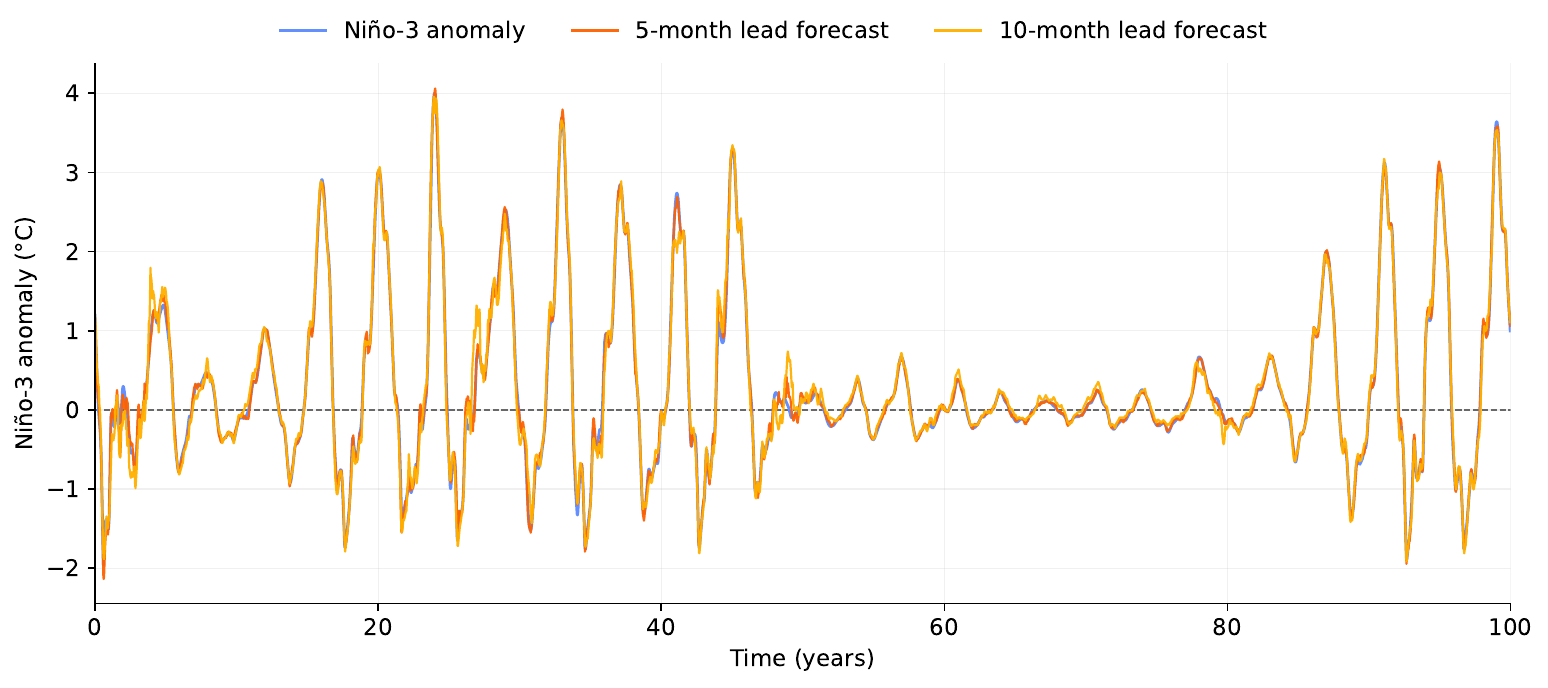}
    \caption{
    The Ni{\~n}o-3 index and CNN forecasts at 5- and 10-month leads over a 100-year segment of the held-out test period.
    The CNN was trained on 10,000 years of ZC data.
    }
    \label{fig:nino3_forecast}
\end{figure*}

We train all three architectures by minimizing the mean-squared error with the AdamW optimizer
\cite{loshchilov2019decoupled}, using a learning rate of $10^{-3}$, weight decay of $10^{-4}$, and a batch size of 256.
We train for at most 100 epochs, stopping when the validation loss fails to improve by at least $10^{-4}$ for 10 epochs, and restore the weights from the best validation epoch.
Figure~\ref{fig:nino3_forecast} illustrates the prediction task using CNN forecasts at lead times of 5 and 10 months.

\section{Results and discussion} \label{sec:results}

This section evaluates the predictive performance of neural networks trained on Zebiak--Cane model data and the quality of the explanations produced by the XAI methods.

\subsection{Data quantity and lead time are more important than architecture for predictive accuracy}

\begin{figure}[t]
\centering
\includegraphics[width=\linewidth]{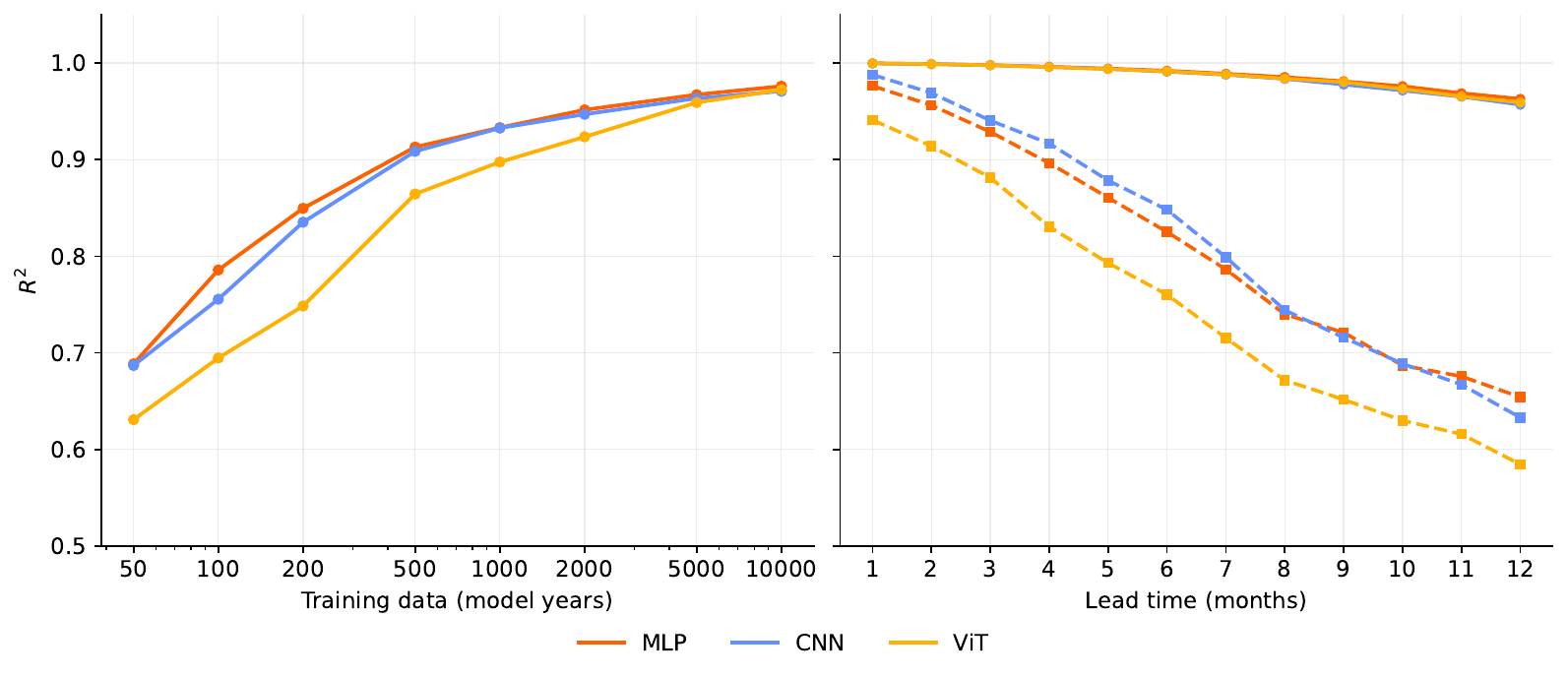}
\caption{
Test-set $R^2$, averaged across five training seeds.
Left: skill at a 10-month lead as a function of training-set size.
Right: skill as a function of lead time for the MLP, CNN, and ViT trained on 10,000 years (solid lines) or 50 years (dashed lines).
}
\label{fig:r2}
\end{figure}

We first evaluate the predictive accuracy of the different neural-network architectures. 
Figure~\ref{fig:r2} shows that accuracy improves with increasing training-set size and decreasing lead time.
With 10,000 years of training data, all three neural networks explain a large fraction of the variance in the Ni{\~n}o-3 index at a 10-month lead time, $R^2 > 0.95$.

Differences among the neural architectures are small but visible in Figure~\ref{fig:r2}.
The MLP and CNN produce the most accurate predictions.
The ViT is generally less accurate than the MLP and CNN given $\leq 2{,}000$ years of training data.
Nevertheless, the differences among neural architectures are much smaller than those produced by variations in lead time and training-set size.

\subsection{AGOP XAI scores highest in three categories}
\label{sec:highest_xai_scores}

We next apply the XAI methods to the models trained on 10,000 years of data when forecasting the most extreme El Ni{\~n}o event in the test set.
We compare four XAI methods:
the gradient (GRAD), Integrated Gradients (IG), GradientSHAP (SHAP), and AGOP XAI (AGOP).
We evaluate the gradient using automatic differentiation at the input state.
We compute the Integrated Gradients attribution vector using a baseline $\bm{\mu} = \bm{0}$,
\begin{equation*}
\bm{\psi}(\bm{x})
=
\bm{x}
\odot
\int_0^1
\nabla f(r\bm{x})
\,\mathrm{d}r,
\end{equation*}
Following \citeA{sundararajan2017axiomatic}, we approximate the integral using a right-endpoint Riemann sum on 1,024 equal subintervals.
We approximate the GradientSHAP attribution by drawing 1,024 reference states $\bm{z}_i$ uniformly without replacement from the training data, drawing independent interpolation coefficients $r_i\sim\operatorname{Uniform}(0,1)$, and then averaging
\begin{equation*}
\frac{1}{1024} \sum_{i=1}^{1024}(\bm{x}-\bm{z}_i)
\odot
\nabla f\bigl(r_i\bm{x}+(1-r_i)\bm{z}_i\bigr).
\end{equation*}
We construct the empirical AGOP using gradients evaluated at all training inputs.
For robustness, we use the nearest $1\%$ of the training inputs.

Applying the XAI methods is fast.
Using gradient batches of size 1,024 on four allocated cores of an AMD EPYC 7662 processor, one CNN explanation required approximately $0.005$ seconds for GRAD, $0.51$ seconds for Integrated Gradients, and $0.48$ seconds for GradientSHAP.
The time to form the AGOP matrix and compute its eigendecomposition was approximately $188$ seconds, after which one AGOP explanation required approximately $0.005$ seconds.

\begin{table*}[t]
\centering
\renewcommand{\arraystretch}{1.2}
\setlength{\tabcolsep}{7.5pt}
\publishedfit{%
\begin{tabular}{l|ccc|ccc|ccc|ccc|}
\hline
 & \multicolumn{3}{c|}{Sensitivity ($S_s$)} & \multicolumn{3}{c|}{Attribution ($S_a$)} & \multicolumn{3}{c|}{Robustness ($S_r$)} & \multicolumn{3}{c|}{Coherence ($S_c$)} \\
\cline{2-4} \cline{5-7} \cline{8-10} \cline{11-13}
 & MLP & CNN & ViT & MLP & CNN & ViT & MLP & CNN & ViT & MLP & CNN & ViT \\
\hline
AGOP & 0.61 & 0.36 & 0.32 & \textbf{0.90} & \textbf{0.83} & \textbf{0.95} & \textbf{0.88} & \textbf{0.86} & \textbf{0.89} & \textbf{0.67} & \textbf{0.80} & \textbf{0.34} \\
GradientSHAP & 0.14 & 0.37 & 0.52 & 0.16 & 0.0015 & 0.60 & 0.58 & 0.55 & 0.64 & 0.38 & 0.48 & 0.13 \\
IG & 0.16 & 0.27 & 0.41 & 0.14 & 0.035 & 0.67 & 0.48 & 0.44 & 0.56 & 0.38 & 0.39 & 0.13 \\
GRAD & \textbf{1.00} & \textbf{1.00} & \textbf{1.00} & 0.87 & 0.20 & 0.037 & 0.64 & 0.41 & 0.46 & 0.59 & 0.51 & 0.13 \\
\hline
\end{tabular}%
}
\caption{
Sensitivity, attribution, robustness, and coherence scores for explanations of the selected extreme El Ni{\~n}o event ($+4.45^\circ\mathrm{C}$) at a 10-month lead.
    The MLP, CNN, and ViT were trained on 10,000 years of ZC data.
Bold indicates the highest score in each column.
}
\label{tab:scores}
\end{table*}

We evaluate the XAI methods using sensitivity, attribution, robustness, and coherence scores defined in Section~\ref{sec:quantification} and present results for the 10-month lead time in Table~\ref{tab:scores}.
In these results, AGOP XAI achieves the highest \textbf{attribution}, \textbf{robustness}, and \textbf{coherence} scores.
GRAD maximizes \textbf{sensitivity} by construction, while AGOP trades sensitivity for stronger performance on the other criteria.

\begin{figure}[!ht]
    \centering
    \includegraphics[
        width=0.95\linewidth
    ]{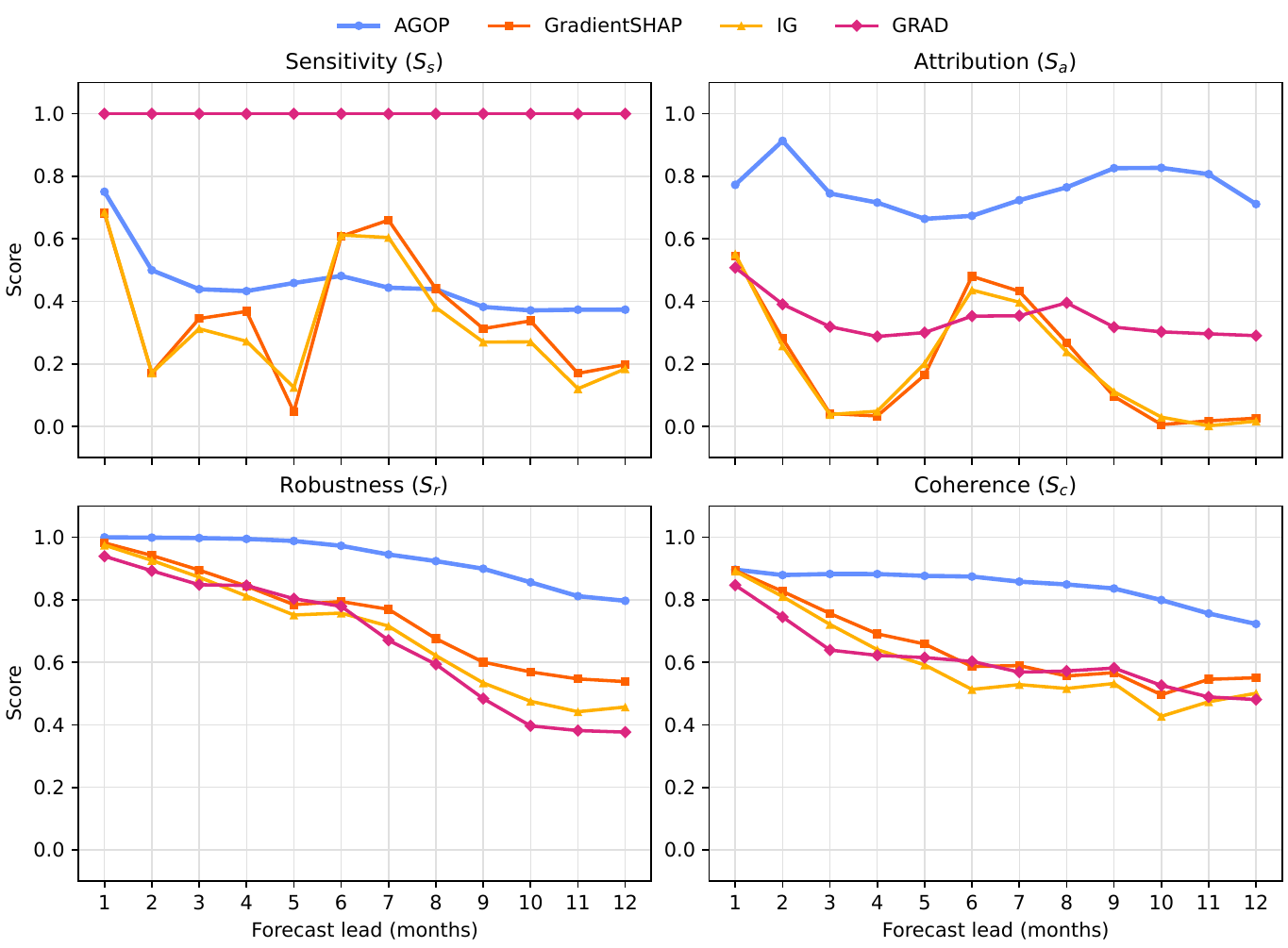}
    \caption{
    Sensitivity, attribution, robustness, and coherence scores for CNN explanations of the selected extreme El Ni{\~n}o event ($+4.45^\circ\mathrm{C}$) at leads from 1 to 12 months.
    Curves show unweighted means across five independently trained CNNs (seeds 42--46); each model used 10,000 years of ZC data.
    }
    \label{fig:cnn_lead_time}
\end{figure}

We next examine how these scores vary with forecast lead in Figure~\ref{fig:cnn_lead_time}.
Several authors \cite{rudin2019stop,bordt2022post,mamalakis2022investigating,bommer2024finding} have identified robustness and coherence issues in post-hoc XAI methods.
At a one-month lead, robustness is $0.94$--$1.00$ and coherence is $0.85$--$0.90$ across the four methods.
At a 12-month lead, baseline robustness and coherence scores fall to $0.38$--$0.55$, whereas AGOP XAI retains a robustness score of $0.80$ and a coherence score of $0.72$.
The advantage of AGOP is therefore greatest at longer lead times.

\begin{figure}[t!]
    \centering
    \includegraphics[width=\linewidth]{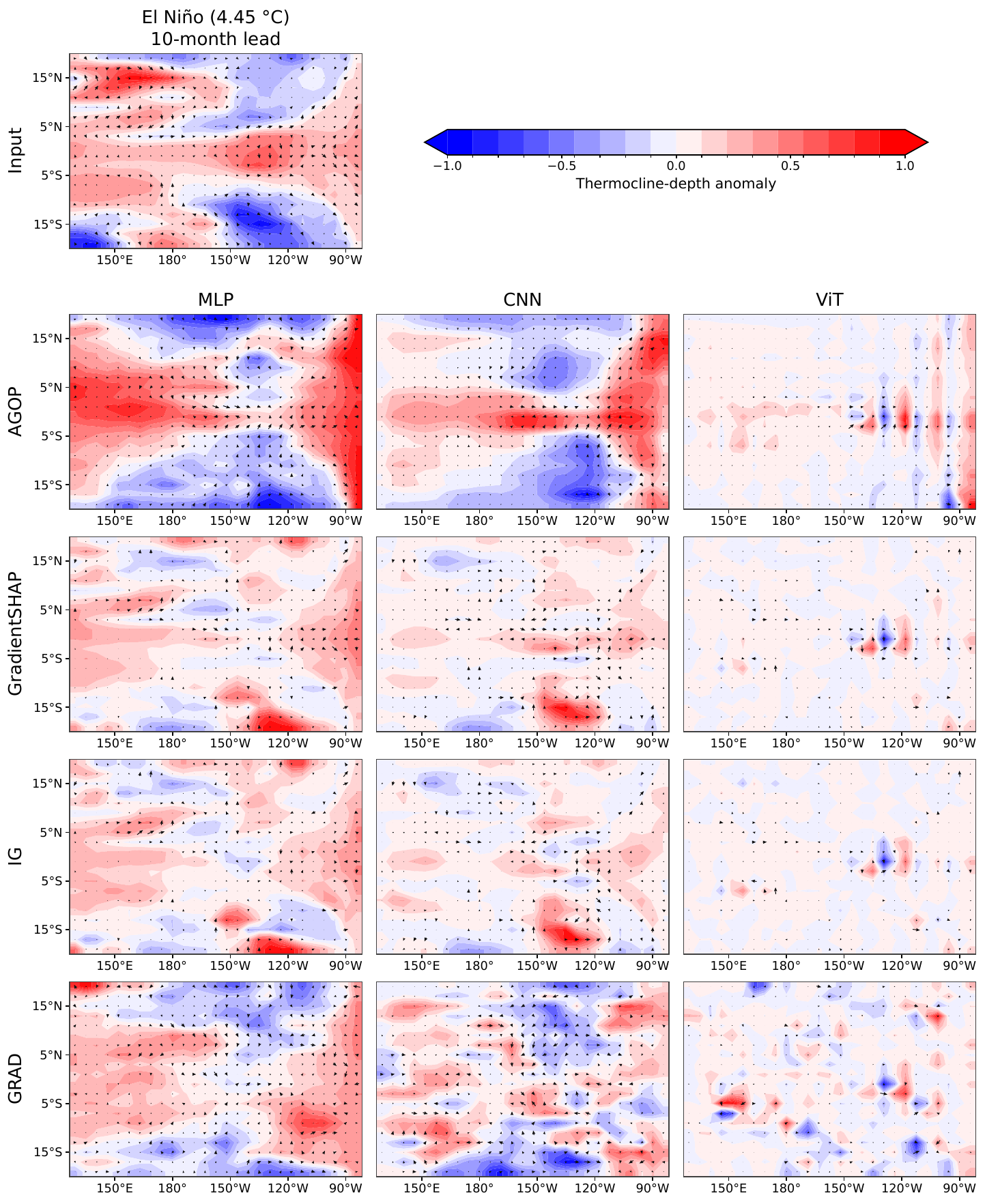}
    
    \caption{
    Top panel: standardized thermocline-depth anomaly (shading) and depth-averaged ocean currents (arrows) 10 months before the selected extreme El Ni{\~n}o event ($+4.45^\circ\mathrm{C}$).
    Remaining panels show AGOP XAI, GradientSHAP, IG, and GRAD explanations (rows) for the MLP, CNN, and ViT (columns).
    Within each panel, shading and current-vector magnitudes are normalized separately to one.
    }
    \label{fig:xai_panels}
\end{figure}

\FloatBarrier

We then visualize the thermocline-depth and ocean-current fields for each XAI method and neural architecture in Figure~\ref{fig:xai_panels}.
At a 10-month lead, these fields together account for $67$--$84\%$ of the sum of squared entries in the AGOP explanation vector across the three architectures.
AGOP XAI produces the smoothest large-scale patterns for all neural architectures, but the coherence depends strongly on the architecture.
The CNN explanation is the most coherent, while the MLP contains small-scale irregularities in the current fields and the ViT thermocline field contains conspicuous patchwise artifacts with no obvious physical interpretation.
The coherence scores reflect these differences: $S_c=0.80$ for the CNN, $S_c=0.67$ for the MLP, and $S_c=0.34$ for the ViT.

\FloatBarrier

\subsection{AGOP explanations reveal physical pathways to El Ni{\~n}o and La Ni{\~n}a}
\label{sec:agop_pathways}

\begin{figure}[t]
\centering
\includegraphics[width=\linewidth]
{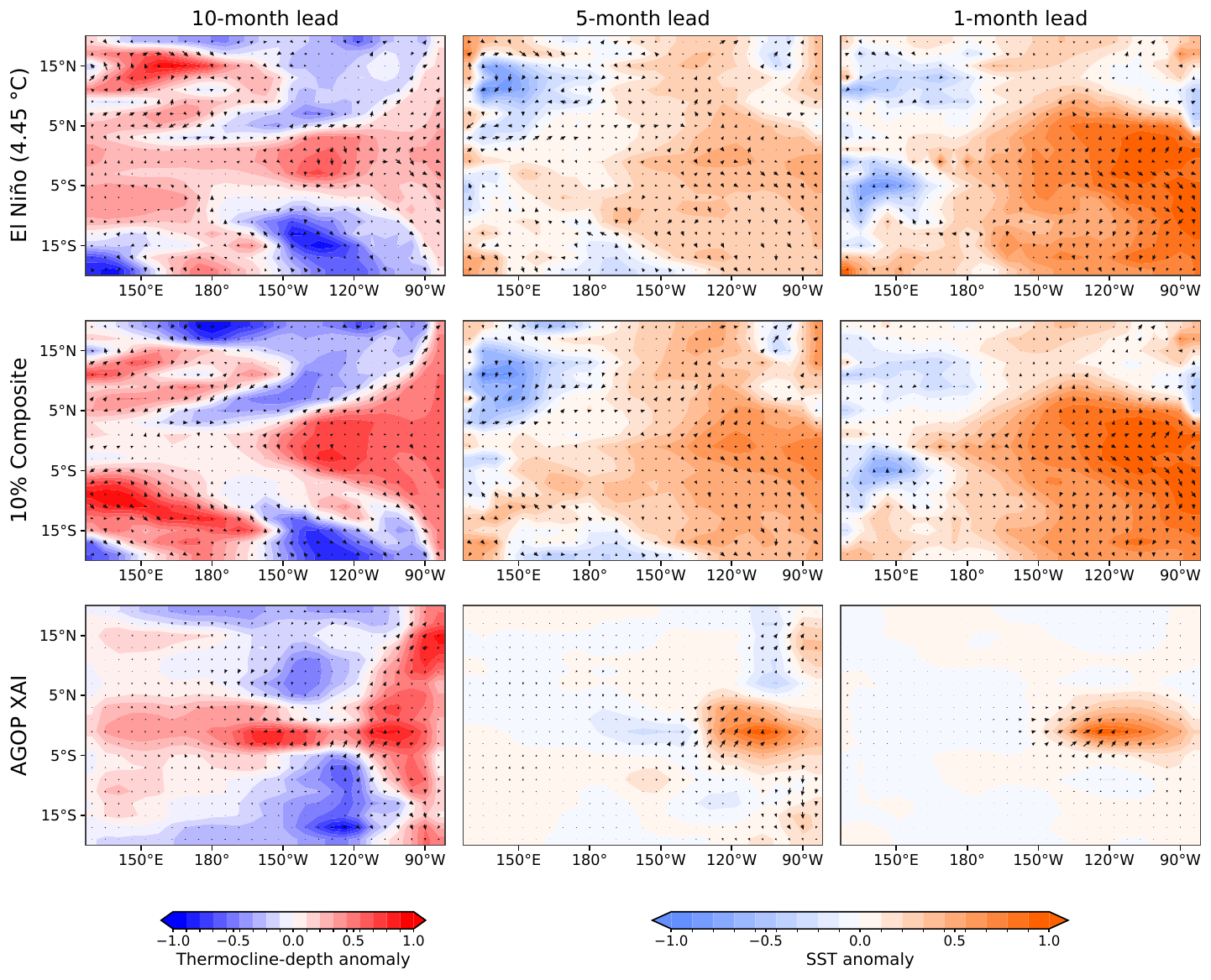}
\caption{
Evolution toward the selected extreme El Ni{\~n}o event ($+4.45^\circ\mathrm{C}$).
Rows show the selected event, a composite of the strongest $10\%$ of complete El Ni{\~n}o episodes in the test set, and the CNN AGOP XAI explanation.
Columns show states 10, 5, and 1 months before the event peak.
Depth-averaged ocean currents are shown at every lead time.
Shading shows thermocline depth at 10 months and SST at 5 and 1 months.
As in Figure~\ref{fig:xai_panels}, shading and arrows are normalized separately within each panel.
}
\label{fig:el_nino}
\end{figure}

\begin{figure}[t]
\centering
\includegraphics[width=\linewidth]
{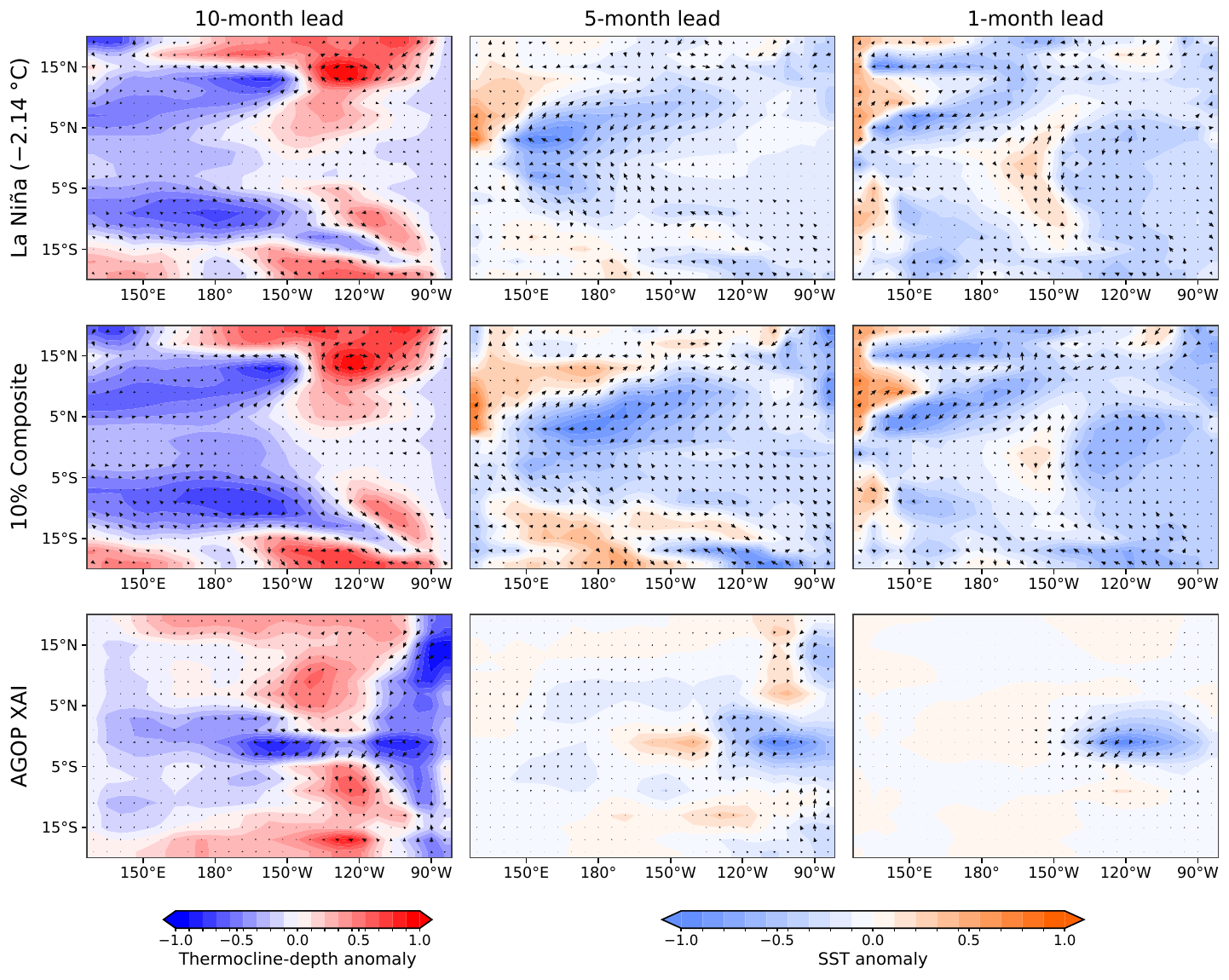}
\caption{
Evolution toward the selected extreme La Ni{\~n}a event ($-2.14^\circ\mathrm{C}$).
Rows show the selected event, a composite of the strongest $10\%$ of complete La Ni{\~n}a episodes in the test set, and the CNN AGOP XAI explanation.
Columns show states 10, 5, and 1 months before the event trough.
Depth-averaged ocean currents are shown at every lead time.
Shading shows thermocline depth at 10 months and SST at 5 and 1 months.
As in Figure~\ref{fig:xai_panels}, shading and arrows are normalized separately within each panel.
}
\label{fig:la_nina}
\end{figure}

We next turn to physical interpretation and examine the evolution toward the selected extreme El Ni{\~n}o ($+4.45^\circ\mathrm{C}$) and La Ni{\~n}a ($-2.14^\circ\mathrm{C}$) events.
Figures~\ref{fig:el_nino} and~\ref{fig:la_nina} compare the selected events, composites of the strongest $10\%$ of events, and the corresponding
CNN AGOP explanations at lead times of 10, 5, and 1 months.
We define complete warm and cold events as contiguous excursions beyond $+1^\circ\mathrm{C}$ and $-1^\circ\mathrm{C}$, respectively, and represent each event by its peak or trough.
Each composite averages the strongest $10\%$ of events, rounded up to 20 of 195 warm events and 23 of 221 cold events.
The selected events and composites show what actually occurred before strong El Ni{\~n}o and La Ni{\~n}a events, whereas AGOP XAI highlights the features that matter most to the CNN forecast.

At a 10-month lead, the El Ni{\~n}o AGOP explanation contains positive thermocline-depth coefficients along the equator and negative coefficients in the southeastern Pacific.
The La Ni{\~n}a thermocline explanation is nearly a sign reversal: the cosine similarity between the two thermocline coefficient fields is $-0.959$.
The equator-to-southeast contrast is prominent in the ZC model data.
Using a basinwide equatorial box ($5^\circ$S--$5^\circ$N, $129.375^\circ$E--$84.375^\circ$W) and a southeastern box chosen to cover the AGOP lobe ($19^\circ$S--$11^\circ$S, $157.5^\circ$W--$95.625^\circ$W), the equatorial anomaly has the expected sign in 19 of 20 warm events and all 23 cold events, while the southeastern anomaly has the expected sign in all 20 warm events and all 23 cold events.

The equator-to-southeast contrast is broadly consistent with recharge--discharge theory, in which equatorial warm-water-volume anomalies precede ENSO SST anomalies \cite{jin1997equatorial,meinen2000warmwater}.
Its Southern Hemisphere asymmetry goes beyond the simplest symmetric recharge picture, although hemispherically asymmetric ENSO-related mass exchange has been documented \cite{kug2003mass}.
This provides precedent for an asymmetric South Pacific contribution but does not validate the precise southeastern lobe found here.
We therefore regard the southeastern lobe as potentially ZC-specific and as a hypothesis for further investigation.

By a five-month lead, SST exceeds thermocline depth in its contribution to the sum of squared entries in the full AGOP explanation.
At a one-month lead, SST accounts for $89.4\%$ of the sum of squared entries in the El Ni{\~n}o explanation and $90.3\%$ in the La Ni{\~n}a explanation.
The short-lead explanations are dominated by warm or cold SST anomalies in the eastern equatorial Pacific.
Unlike the broad SST anomalies for the extreme events and composites, the AGOP anomalies are spatially localized in the eastern Pacific, highlighting the areas that matter most to the CNN forecast.
Together, these results suggest a transition from subsurface preconditioning at long leads to SST persistence near the event extremum.

The ocean-current patterns are best interpreted as CNN forecast sensitivities.
At a 10-month lead, the El Ni{\~n}o explanation contains positive zonal-current coefficients near the equator and equatorward meridional-current coefficients, while the La Ni{\~n}a pattern is approximately reversed.
The two current fields together account for $54.5\%$ of the sum of squared entries in the El Ni{\~n}o AGOP explanation and $58.0\%$ in the La Ni{\~n}a explanation.
However, their cosine similarities with the corresponding composites are only $0.33$ and $0.36$, respectively.
Thus, the currents represent strong CNN sensitivity directions but are not typical features of the extreme-event composites.

\FloatBarrier

\subsection{AGOP explanations can steer extreme ENSO behavior in the physical model}
\label{sec:agop_steering}

Finally, we test whether directions extracted from CNN explanations can steer the ZC model itself.
Success would provide additional evidence that AGOP XAI identifies physically relevant features.

We focus on AGOP explanations of the 10-month-lead CNN forecasts for the selected extreme El Ni{\~n}o and La Ni{\~n}a events.
Each CNN input, and hence each AGOP explanation, consists of two annual-phase coordinates and 2,160 spatial values from four fields on the $20\times27$ grid.
We let $\bm{e}$ denote the unit-normalized spatial AGOP direction.
Our intervention vector contains 28,591 coordinates in five native ZC arrays representing Kelvin-wave amplitude, non-Kelvin zonal current, meridional current, thermocline depth, and SST anomaly.
Because the two representations use different variables and grids, we cannot simply add a multiple of $\bm{e}$ to a ZC checkpoint.

Instead, we apply nine native-state perturbations $\bm a_1,\ldots,\bm a_9$ during the three months preceding the release point for a 10-month forecast.
We optimize the perturbations to approximately solve
\begin{equation*}
\underset{\bm a_1,\ldots,\bm a_9}{\operatorname{minimize}}
\quad
\sum_{i=1}^{9}
\bm a_i^\top \bm C^+ \bm a_i
\qquad
\text{subject to}
\qquad
\bm e^\top \bm x_{\mathrm{opt}}
=
q,
\end{equation*}
where $\bm x_{\mathrm{opt}}$ is the standardized spatial CNN input at the end of the intervention window and $q$ is the desired AGOP coordinate.
We form $\bm C$ by averaging over all 360,000 training-period states after subtracting the corresponding phase-specific means.
Here, $\bm C^+$ denotes the Moore--Penrose pseudoinverse.
We restrict each perturbation to the range of $\bm C$, so variations commonly observed in the ZC simulation are relatively inexpensive, while unusual combinations of variables are expensive.
We solve the constrained problem using exact adjoint derivatives and nonlinear sequential quadratic programming.
For difficult targets, we approach the requested AGOP coordinate in stages.

\begin{figure}[!ht]
    \centering
    \includegraphics[width=\linewidth]
    {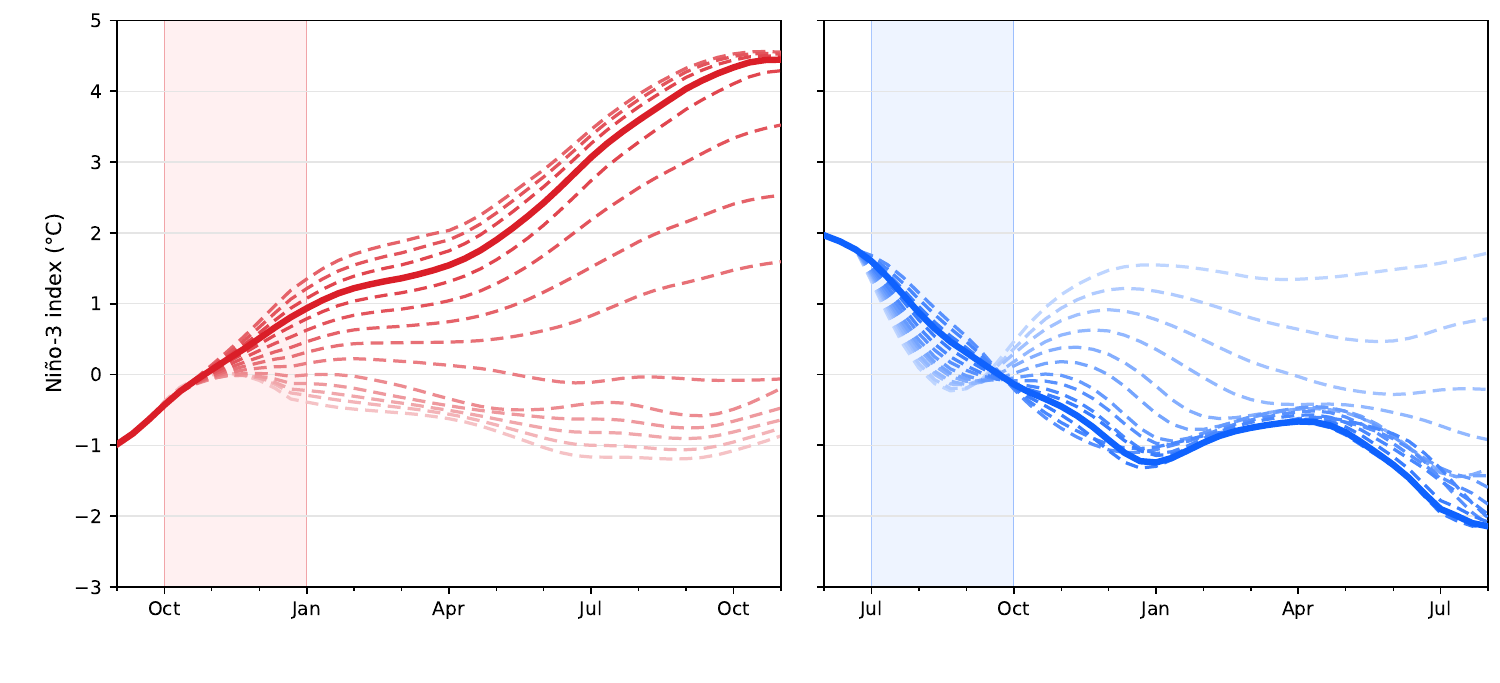}
    \caption{
    AGOP dose--response experiments for the selected extreme El Ni{\~n}o ($+4.45^\circ\mathrm{C}$; left) and La Ni{\~n}a ($-2.14^\circ\mathrm{C}$; right).
    Solid curves show the unnudged events ($c=1$).
    Dashed curves show trajectories after nine optimized native-state perturbations change the release-state AGOP coordinate from $q_{\mathrm{ext}}$ to $c q_{\mathrm{ext}}$; curves become lighter as $|c-1|$ increases.
    Shading marks the three-month intervention window, followed by ten months of free ZC evolution.
    }
    \label{fig:stronger_weaker}
\end{figure}

In the first test, which is analogous to the sensitivity metric, we begin with the selected extreme events and change their AGOP coordinate from
$q_{\mathrm{ext}}$
to $c\, q_{\mathrm{ext}}$, where
\begin{equation*}
c \in \{0.0, 0.1, \ldots, 0.9, 1.1, 1.2, 1.3\}.
\end{equation*}
The unnudged event corresponds to $c = 1$, and each listed value requires a separate nonlinear optimization.
As shown in Figure~\ref{fig:stronger_weaker}, multipliers above one produce little or no additional intensification, consistent with saturation or nonlinear reversion.
Reducing the AGOP coordinate eventually weakens and reverses both events, demonstrating that the extreme events are sensitive to the AGOP pattern.

In the second test, which is analogous to the attribution metric, we ask whether the AGOP coordinate can steer the ZC model toward strong warm and cold events from non-extreme ZC initial states.
We sample ten nonoverlapping test-set trajectories uniformly without replacement from states having the appropriate annual phase.
We use the same underlying trajectories for both events, shifting the release points by three months to match the phases of the selected extrema.
For each release point, we seek perturbations that raise its event-oriented AGOP coordinate to the value attained by the corresponding selected extreme event.

\begin{figure}[t]
    \centering
    \includegraphics[width=\linewidth]
    {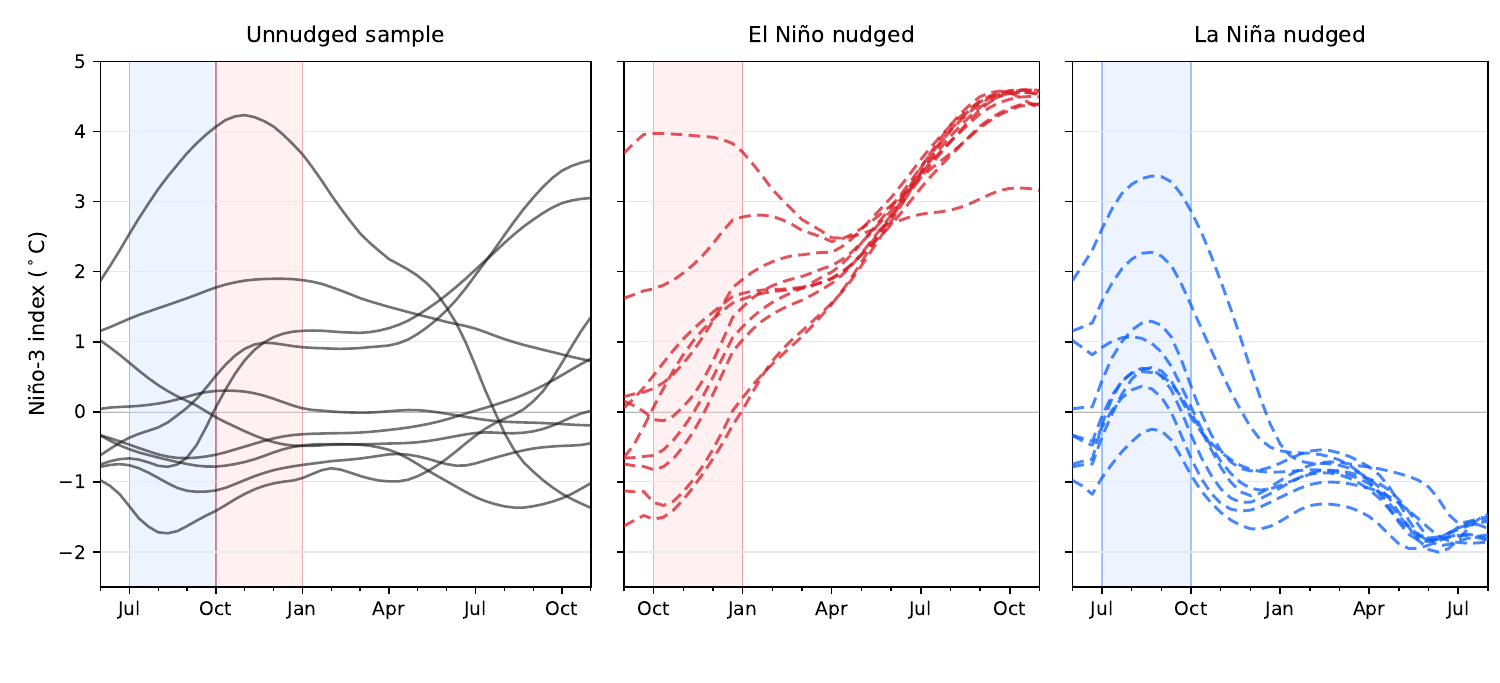}
    \caption{
    AGOP steering experiments for ten uniformly sampled test trajectories.
    The left panel shows the shared unnudged trajectories.
    The middle and right panels show the ten completed El Ni{\~n}o and nine completed La Ni{\~n}a experiments, respectively.
    Solid curves are unnudged, and dashed curves are AGOP-steered trajectories.
    Shading marks each three-month intervention window, followed by ten months of free ZC evolution.
    }
    \label{fig:sampling}
\end{figure}

Figure~\ref{fig:sampling} shows that nine of the ten El Ni{\~n}o experiments terminate between $4.33^\circ\mathrm{C}$ and $4.59^\circ\mathrm{C}$, close to the reference value of $4.45^\circ\mathrm{C}$; the remaining experiment reaches $3.16^\circ\mathrm{C}$.
All nine completed La Ni{\~n}a experiments terminate between $-1.86^\circ\mathrm{C}$ and $-1.46^\circ\mathrm{C}$, compared with the $-2.14^\circ\mathrm{C}$ reference.
One La Ni{\~n}a optimization did not satisfy the prespecified numerical acceptance criteria even after adaptive continuation, so it is excluded.

\begin{figure}[t]
    \centering
    \includegraphics[width=\textwidth]{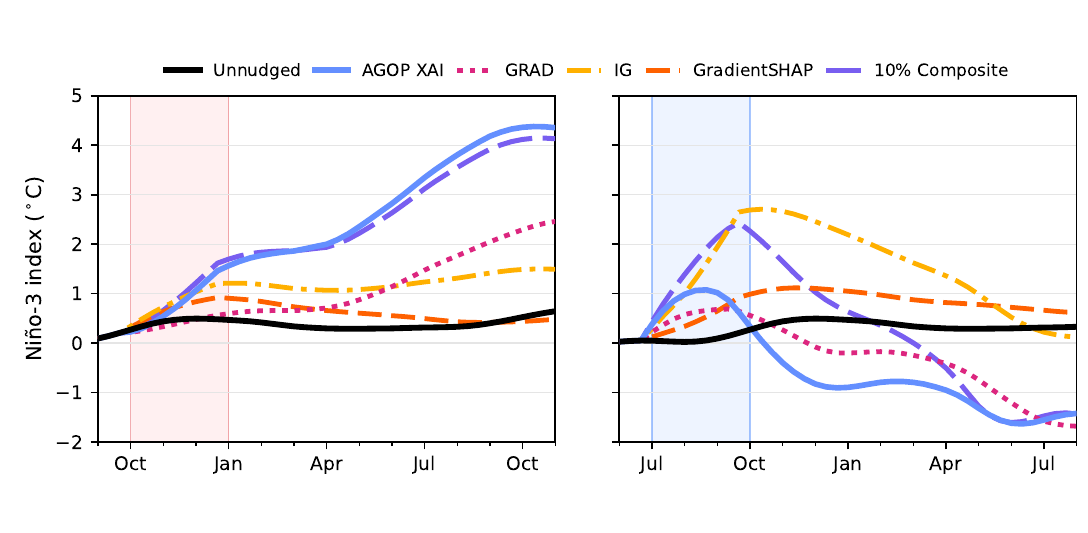}
    \caption{
    Mean Ni{\~n}o-3 trajectories for the El Ni{\~n}o (left) and La Ni{\~n}a (right) steering experiments.
    Each colored curve adds the mean response among completed optimizations to the common 10-member control mean; shading marks the three-month intervention window.
    The number of completed experiments is $n=10$ for El Ni{\~n}o.
    For La Ni{\~n}a, it is $n=9$ for AGOP, $n=8$ for GRAD, $n=6$ for IG, and $n=10$ for GradientSHAP and the $10\%$ composite.
    }
    \label{fig:nudging}
\end{figure}

For comparison, Figure~\ref{fig:nudging} summarizes steering experiments using directions extracted from GRAD, IG, GradientSHAP, and the $10\%$ composites.
AGOP produces the largest mean warming in the El Ni{\~n}o experiment.
In the La Ni{\~n}a experiment, AGOP, GRAD, and the composite produce similarly strong mean cooling.
Every completed AGOP, composite, and GRAD experiment changes the terminal Ni{\~n}o-3 index in the intended direction, whereas IG and GradientSHAP are less consistent.

Taken together, these experiments show that the AGOP direction is not merely a description of CNN sensitivity.
It provides the most effective overall steering among the XAI directions tested in the physical ZC model.

\FloatBarrier
\section{Conclusions} \label{sec:conclusions}

We introduced four scores for evaluating AI explanations: sensitivity, attribution, robustness, and coherence.
We also introduced AGOP XAI, which combines gradients across the training data to identify input-specific patterns in physical fields.
Together, these tools supplement standard measures of predictive skill with quantitative tests of explanation quality.

Across the architecture and lead-time comparisons reported here, AGOP XAI provides the strongest overall performance.
The gradient maximizes sensitivity by construction, whereas AGOP XAI generally achieves the highest attribution, robustness, and coherence scores.
Its advantage is greatest at long lead times, where the baseline explanations become less robust and spatially coherent.

For the selected 10-month CNN forecasts, AGOP XAI identifies an equator-to-southeastern-Pacific thermocline contrast associated with both extreme El Ni{\~n}o and La Ni{\~n}a events.
At shorter leads, the explanations shift toward persistent SST anomalies in the eastern equatorial Pacific.
Steering uniformly sampled, phase-matched ZC trajectories along the corresponding 10-month AGOP explanation generally produces an intense event of the intended sign 10 months later.
Reducing the AGOP coordinate weakens and, for sufficiently large reductions, reverses the selected extreme events.
These steering results support the physical relevance of the identified pattern.

\section*{Acknowledgments}

D.S.A. is supported in part by the U.S. National Science Foundation under Award No.~RISE-2425898.
R.J.W. is supported by a 2026--2027 Hellman Fellowship.
We thank Eli Tziperman for discussions about the Zebiak--Cane model and how to interpret it, Libby Barnes for 
explaining the difference between sensitivity and attribution to us, and Kerry Emanuel for proposing that we test the XAI explanation vectors in the physical model. 

During the preparation of this work, the authors used ChatGPT to develop code, generate figures, tables, and reproducibility artifacts, and improve manuscript readability.
The authors reviewed, tested, and substantially revised the AI-assisted output and take full responsibility for the content of the published article.

\section*{Conflict of Interest}

The authors declare no conflicts of interest.

\section*{Open Research Statement}
The software used for data generation, neural-network training, XAI, steering, and figure production is available from \url{https://github.com/rjwebber/agop-xai/releases/tag/v1.0.0} under the BSD 3-Clause License \cite{hui2026agopxai_software}.
The processed ZC simulation, trained models, and reproducibility artifacts are available from \url{https://doi.org/10.5281/zenodo.23053004} under CC BY 4.0 \cite{hui2026agopxai_collection}.

\bibliography{refs}

\end{document}